\documentclass[aps,pre,twocolumn]{revtex4}
\pdfoutput=1 
\usepackage{graphicx}
\usepackage{amssymb}
\usepackage{amsmath}
\usepackage{color}
\usepackage{epstopdf}
\usepackage{bm}
\usepackage[normalem]{ulem}

\DeclareMathOperator\erf{erf}

\newcommand{\beq}{\begin{equation}}
\newcommand{\eeq}{\end{equation}}
\renewcommand{\footnoterule}{%
  \kern -3pt
  \hrule width \columnwidth height 1pt
  \kern 2pt
}

\begin{document}

\title{Origin and scaling of chaos in weakly coupled phase oscillators}

\author{Mallory Carlu, Francesco Ginelli, and Antonio Politi}
\address{SUPA, Institute for Complex Systems and Mathematical Biology, King’s College, University of Aberdeen,
Aberdeen AB24 3UE, United Kingdom}

\date{\today}

\begin{abstract}
We discuss the behavior of the largest Lyapunov exponent $\lambda$ in the incoherent phase 
of large ensembles of heterogeneous, globally-coupled, phase oscillators. 
We show that the scaling with the system size $N$ depends on the details of the
spacing distribution of the oscillator frequencies. 
For sufficiently regular distributions $\lambda \sim 1/N$, while for strong fluctuations
of the frequency spacing, $\lambda \sim \ln N/N$ (the standard setup of
independent identically distributed variables belongs to the latter class).
In spite of the coupling being small for large $N$, the development of a rigorous perturbative theory
is not obvious. In fact, our analysis relies on a combination of various types of 
numerical simulations together with approximate analytical arguments, based on a suitable
stochastic approximation for the tangent space evolution.
In fact, the very reason for $\lambda$ being strictly larger than zero is the presence
of finite size fluctuations. We trace back the origin of the logarithmic correction to a weak
synchronization between tangent and phase space dynamics.
\end{abstract}

\maketitle

\section{Introduction}

Oscillatory phenomena are ubiquitous in nature. Collections of interacting oscillators 
have been widely employed to model a large number of phenomena, 
ranging from biological \cite{Kbio1, Kbio2, Kbio3}, to social \cite{Ksoc},
and chemical \cite{Kchem1, Kchem2} oscillators. Physical examples include 
lasers \cite{KLaser} and arrays of non-identical Josephson junctions \cite{WCS96}.
While dealing with oscillator ensembles, much attention has been of course devoted to the phenomenon of
collective synchronization \cite{Synch}, in which a (large) system of heterogeneous oscillators 
spontaneously locks at a common frequency above some critical coupling strength \cite{Winfree}.

In the weak-coupling limit the oscillators are characterized by a single variable, the phase,
while the mutual interaction is mediated by a coupling function which depends
only on mutual phase differences (this is the so-called Kuramoto-Daido model~\cite{Daido1}).
As a result, an ensemble of $N$ oscillators is characterized by at most $N-1$ effective degrees of freedom.

In the thermodynamic limit $N\to\infty$, the evolution is regular both below and above the 
synchronization transition, and the largest Lyapunov exponent (LLE) equal to zero
(the opposite claims contained in Ref.~\cite{Rapisarda} are probably due to errors in the numerical simulations).
In the small-$N$ limit, the minimal number of oscillators to observe chaos is 4. 
This bound has a simple justification: it is, in fact, well known that it is necessary
to have at least three (independent) variables to generate chaos.
Less obvious is that while $N=4$ is sufficient in heterogeneous oscillators~\cite{Popovych}, the same
bound is attained for identical oscillators only if the coupling function is not purely sinusoidal~\cite{Ashwin},
i.e. going beyond the Kuramoto model~\cite{Kuramoto75, Kuramoto84a, Kuramoto84b, Kuramoto86}.

For large but finite $N$, there is clear evidence that the LLE is strictly positive~\cite{Popovych}, but little is
known about the underlying mechanisms, even though one might expect that a perturbative approach can be set in
the thermodynamic limit, when the effective coupling becomes increasingly weak. 
In this paper, with the help of direct numerical simulations and approximate analytical arguments, 
we discuss the finite-size scaling of the LLE and the origin of chaos in the incoherent phase of 
{\it globally coupled} heterogeneous phase oscillators. More specifically, we focus our attention on the 
celebrated Kuramoto model, the simplest example of heterogeneous, globally coupled, phase oscillators. 
Its dynamics reads

\begin{equation}
\dot{\theta}_i = \omega_i + \frac{g}{N} \sum_{j=1}^N \sin \left(\theta_j -
  \theta_i \right)
\label{Kuramoto}
\end{equation}
where $g$ is the global coupling parameter, $\theta_i$ the phase of oscillator $i$ (with
$i=1,2,\ldots,N$), 
and the $\omega_i$ are the quenched natural frequencies, typically
drawn out of some distribution $P(\omega)$. 

Numerical simulations give compelling evidence that the LLE converges to zero in the
thermodynamic limit. However, it may scale into two different ways,
\begin{equation}
\lambda (N)  \sim \frac{\ln N}{N} \, , \qquad \mathrm{(class\ I)}
\label{log-scaling}
\end{equation}
or
\begin{equation}
\lambda (N) \sim \frac{1}{N} \, , \qquad \mathrm{(class\ II)}
\label{norm-scaling}
\end{equation}
While we are not able to identify exactly the necessary and sufficient conditions for a
distribution to belong to a given class, we can at least safely
state that whenever the frequencies are generated independently of each other
(irrespective whether the distribution itself is Gaussian, uniform in
a finite interval, or other), then
the logarithmic correction is always present (class I). Independently
generated frequencies are of course the most natural choice for many physical or biological setups.
On the contrary, if the finite set of frequency spacings is selected
in a {\it regular} fashion so as to follow the ``macroscopic" shape
$P(\omega)$ down to the tiniest scales, the LLE scales as in~(\ref{norm-scaling}).
This is for instance the case of strictly equispaced frequencies (as
already shown in Ref. \cite{Popovych}), a set-up which is sometimes
considered to be a typical representative of a uniform distribution to avoid the burden of
averaging over different realizations of the frequencies.

In practice our analysis reveals that the value of the LLE depends strongly on tiny 
differences in the actual values of the frequencies, which disappear when the
distribution is coarse-grained.
The presence of quasi-degeneracies (almost identical frequencies) definitely enhances the LLE.
In fact, as we see in the next section, this is a major source of differences in the
typical values of the LLE. However, the overall scenario is more complex than that; the
scaling behavior of the LLE depends also on other 
details such as the presence of long-range order.

Anyway, the analytical arguments based on a (self-consistent) stochastic
approximation of the chaotic dynamics show that the very existence of chaos 
is due to fluctuations in the tangent space dynamics. Finally, we also put
forward a conjecture that traces back the origin of the logarithmic
correction of class I to a sort of weak synchronization
between tangent- and phase-space coordinates.

The remainder of this paper is organized as follows. In Section II, we
briefly review the Kuramoto model and present the outcome of numerical
simulations (in the incoherent state) performed for different frequency distributions.
In Section III, we make use of simplified quasi-periodic and
discrete-time approximations to put the conjectured existence of two different
universality classes on more solid grounds.
Semi-analytical arguments, based on a stochastic
approximation, are put forward in Section IV.
Finally, our results are discussed in the concluding Section. 

\section{The Kuramoto model}

Since its introduction more than 40 years ago, the Kuramoto model
(\ref{Kuramoto}) has attracted a good deal of attention in the
scientific literature and it has been studied with a
combination of analytical techniques and numerical approaches
\cite{Kreview1}. 

In particular, it is well known \cite{Kreview1} that for a
symmetric and unimodal distribution $P(\omega)$, a phase
transition is observed for
$g = g_c \equiv 2/(\pi\,P(\bar{\omega}))$, where
$\bar{\omega} \equiv \int \omega P(\omega) d\omega\;$
is the mean frequency~\cite{NOTE2}. 
Below $g_c$ all oscillators are effectively uncoupled and their phases uniformly distributed;
above $g_c$ we are in the presence of a symmetry broken, {\it partially synchronized} phase, 
characterized by a finite (complex) order parameter
\begin{equation}
R \mathrm{e}^{i \psi} = \frac{1}{N} \sum_{j=1}^N \mathrm{e}^{i \theta_j} \,.
\label{OP}
\end{equation}
In the partially synchronized phase, a finite fraction of
oscillators lock at the mean frequency $\bar{\omega}$.
They are those whose natural frequency lies closer to the mean one,
i.e. in the most densely distributed region for a symmetric unimodal distribution.

The transition is different for non-unimodal distributions. 
In particular, for the degenerate case of a uniform $P(\omega)$
on a compact interval, one has a first order transition
between a standard incoherent phase and a fully synchronous state, where all
oscillators are locked to the same linearly stable periodic orbit with 
$\dot{\theta}_i = \bar{\omega}$ \cite{Tanaka, vanHemmen}.

Finite systems are notoriously harder to analyze. 
Advances have been made concerning, for instance, corrections to the 
thermodynamic critical coupling parameter value \cite{Strogatz2016},
and in the analysis of the dynamical fluctuations of the order parameter~\cite{Daido}.
As mentioned in the previous section, numerical studies of large but finite systems are
particularly challenging, due to the quenched noise: the results of simulations,
in order to be representative of a specific size,
have to be averaged over many different realizations of the
natural frequencies drawn out of the selected distribution $P(\omega)$. 
This can be particularly demanding if the observable of interest decreases with
the system size (at it is the case of the LLE) and keeps being affected by
strong sample-to-sample fluctuations.

In order to get rid of the problem of averaging, some researchers choose to
study a single realization of $P(\omega)$, under the assumption 
that the microscopic details do not alter the scaling behavior.
The selection is typically made by choosing the single frequencies
so that the corresponding cumulative probability densities are equispaced
(see the following subsection for a precise definition).
We are going to follow also this approach in the case of both uniform
and Gaussian distributions.
This strategy greatly reduces the computational burden, but, as already
anticipated, we find that the LLE scales differently in these regular cases.
In the attempt of clarifying the origin of the different scaling, we introduce
and analyse further setups with different degree of (long-range) order and
different distributions of the spacing between consecutive frequencies $\omega_j$.
We will see that these ingredients do not only contribute to the quantitative value
of the LLE, but also to its scaling behavior with $N$. 

It is often instructive to rewrite the Kuramoto model to explicitate
the mean field nature of the global coupling. 
Making use of the order parameter expression (\ref{OP}), one can immediatly rewrite the
Kuramoto dynamics (\ref{Kuramoto}) as
\begin{equation}
\dot{\theta}_i = \omega_i + g\, R \sin \left(\psi -
  \theta_i \right)\,,
\label{Kuramoto2}
\end{equation}
revealing that the dynamics of each oscillator is ruled by its phase
difference with the average phase $\psi$ and by the interaction strength $gR$.

In the thermodynamic limit, the order parameter $R \mathrm{e}^{i \psi}$
settles to either a fixed point $R=0$, or to a limit cycle (a uniform rotation
at the average frequency $\bar{\omega}$).
In finite systems, however, the order parameter is characterized
by fluctuations. In the incoherent regime, for instance \cite{Daido}
$R \sim \frac{1}{\sqrt{N}}$. 
These fluctuations are the ultimate source of a strictly positive LLE \cite{NOTE1}.

\subsection{The different frequency setups}
\label{sec:classes}

The very same distribution $P(\omega)$ can be generated in different ways. One can see this in
the following way. On the one hand, let us order the
frequencies $\omega_i$ from the smallest to the largest one to produce $y\equiv i/N$ versus $\omega_i$. 
On the other hand, consider the cumulative distribution function
\begin{equation}
\Phi(\omega) \equiv \int_{-\infty}^{\omega_j} P(\omega) d\omega \; .
\label{eq:cumul}
\end{equation} 
So long as in the large-$N$ limit, $y(\omega_i)$ converges to $\Phi(\omega)$, one
can claim that the $N$ frequencies $\omega_i$ are distributed
according to $P(\omega)$. However, for finite $N$, different choices
can be made, leading to different finite size setups. 
In the following, we consider various choices. Here we introduce and
analyse the most natural {\it disordered} and {\it regular}
setups; further options are introduced and discussed in the next section.

\begin{itemize}
\item {\it Disordered, Gaussian (DG) distributed frequencies}: the frequencies
are distributed independently of each other and drawn from
a normal distribution -- perhaps the most popular choice --
with zero average and unit standard deviation,
\begin{equation}
P_G(\omega) = \frac{1}{\sqrt{2\pi}} \mathrm{e} ^{-\frac{\omega^2}{2}}
\label{gauss}
\end{equation}
Some preliminary studies can be found in~\cite{PikovPoliti}, where a slower than $1/N$ is
reported.

\item {\it Regular, Gaussian (RG) set of frequencies}: given a generic
distribution $P(\omega)$, $N$ frequencies $\{\omega_j\}$ ($j=1,2,\ldots,N$) are
generated as follows
\begin{equation}
\frac{j-0.5}{N} = \Phi (\omega_j)  \; ,
\label{eq::detdistrib}
\end{equation} 
where the $1/(2N)$ shift is introduced to make the definition more symmetric.
In the case of a Gaussian distribution (\ref{gauss}), 
$\erf(z)=2\Phi(z\sqrt{2})-1$, where $\erf$ is the error function.
In practice, the frequencies $\omega_j$ are determined by applying a Newton's method to find the zeros of 
\begin{equation}
h(\omega_j)\equiv \frac{j-0.5}{N} - \frac{1}{2} \left(1+ \erf\left(\frac{\omega}{\sqrt{2}}\right)\right) \,.
\label{eq::detdistrib2}
\end{equation} 
This distribution minimizes the fluctuations of the frequency spacing. It has been already used to study
critical properties of the Kuramoto model~\cite{Chate2015} 
In the thermodynamic limit, the synchronization transition
takes place at $g_c=\sqrt{8/\pi}$ for both DG and RG sets of frequencies \cite{Kreview1}.

\item {\it Disorederd, uniformly (DU) distributed frequencies}: as for DG,
the frequencies are uniformly distributed id variables,
\begin{equation}
P(\omega) = \left\{
\begin{array}{c c}
1 &\mbox{for}\;\omega \in [-1/2,1/2]\\
0 & \mbox{otherwise}
\end{array}\right.
\label{uniform}
\end{equation}
As already noted, this set up shows a first-order phase transition from an
incoherent to a fully phase-locked phase at $g_t= 2 / \pi$.

\item {\it Regular, equispaced (RE) set of frequencies}:
Applying Eq.~(\ref{eq::detdistrib}) with the uniform distribution
yields a set of equally spaced frequencies
\begin{equation}
\omega_j = \frac{2j-1}{2N} - \frac{1}{2}
\label{eq::eqspaced}
\end{equation}
This setup has been repeatedly investigated~\cite{vanHemmen, Popovych, Pazo2005} as 
a testing ground for the properties of the Kuramoto model.
\end{itemize}

Before
proceeding, we remark that, in the following, without loss of generality, we will work in a
uniformly rotating frame so that $\bar{\omega}=0$ and the natural
frequency distribution $P(\omega)$ is symmetric around
zero. Frequencies will be ordered from the smallest to largest one.

In the disordered setups we expect the LLE $\lambda_{\alpha}$
to depend on the realization $\alpha$ of the stochastic process, so
that it is necessary to average over a sufficiently large set of
realizations.  The resulting ensemble average 
\begin{equation}
\lambda  = \langle \lambda_{\alpha} \rangle_\Omega \, ,
\label{LEdis}
\end{equation}
is our first object of investigation, where $\Omega$ denotes the cardinality
of the set. In the regular setups no averaging is obviously requried.

\subsection{Finite size scaling of the largest Lyapunov exponent}
\label{sec:param}

We begin our journey by performing a numerical finite-size scaling analysis of the
LLE for the Kuramoto model~(\ref{Kuramoto}). Tangent-space dynamics is ruled by the equation
\begin{equation}
\dot{\delta \theta}_i = 
\frac{g}{N} \sum_j  \cos (\theta_j - \theta_i) (\delta \theta_j - \delta \theta_i) =
 \sum_{j=1}^N J_{i j} ({\bm \theta}) \, \delta \theta_j  \, ,
\label{KuramotoTS}
\end{equation}
where the upper dot denotes the time derivative, while the Jacobian matrix 
\begin{equation}
J_{ij} ({\bm \theta}) = \frac{g}{N} \left( \cos (\theta_j - \theta_i) -  \delta_{ij} \,\sum_k \cos (\theta_k - \theta_i) \right)
\label{Jacobian}
\end{equation}
is evaluated along the phase-space trajectory ${\bm \theta}(t) \equiv \left(
\theta_1, \theta_2, \dots, \theta_N \right)$. The symbol $\delta_{ij}$
denotes as usual the Kronecker's delta.
In practice the matrix $J_{ij}$ is the sum of a full and a diagonal matrix such that the sum of all elements along 
each row is zero. In other words this is a typical instance of (extended) Laplacian coupling.

It should be noticed that the parameter $g$ cannot be scaled out in spite of it being just
a multiplicative factor. In fact, while $g$ can be explicitly removed by rescaling the time variable,
the dependence on $g$ would be transferred to the time evolution of the angles $\theta_i$.
In practice, controlling $g$ is equivalent to controlling the frequency dispersion.

It is convenient to rewrite the evolution equation as
\begin{equation}
\dot{\delta \theta}_i = 
g [-R \cos(\psi-\theta_i) \delta \theta_i + Z \cos (\beta- \theta_i)]
\label{KTS2}
\end{equation}
where $R$ is the Kuramoto order parameter and $\psi$ is its phase, while
\begin{equation}
Z \mathrm{e}^{i\beta} = \frac{1}{N} \sum_j  \delta \theta_j
\mathrm{e}^{i\theta_j} 
\label{eq:wa}
\end{equation}
is the average of the tangent-space variables $\delta \theta_j$ oriented according to the corresponding oscillators phases. 
We shall discuss the role of this term more in detail in Section~\ref{theo}.

In the following, we integrate the Eqs.~(\ref{Kuramoto2},\ref{KTS2}) using a 4th order Runge-Kutta
algorithm, with a time step $\delta t = 0.01$. The LLE
$\lambda$ measures
the asymptotic growth of a generic tangent-space vector ${\bm
  \delta\theta}(t) \equiv \left( 
\delta\theta_1, \delta\theta_2, \dots, \delta\theta_N \right)$,
\begin{equation}
\lambda = \lim_{t \to \infty} \frac{1}{t} \ln \frac{\|{\bm \delta\theta}(t) \|}{\|{\bm \delta\theta}(0) \|}\,.
\label{LE1}
\end{equation}
In practice, the tangent-space vector is rescaled to unit norm 
every $\Delta t$ time units (unless otherwise stated, we choose $\Delta t = 1$). 
This procedure allows us to  reconstruct the finite-time Lyapunov
exponent (FTLE) as
\begin{equation}
\lambda_t = \frac{\ln \; \|{\bm \delta \theta}\|}{\Delta t} \, ,
\label{FTLE}
\end{equation}
where $\|{\bm \delta\theta} \|$ is the norm of the tangent vector immediately
before rescaling. The LLE of the dynamics is then the
asymptotic time average of $\lambda_t$.

As anticipated, we concentrate on the incoherent phase. For the Gaussian
setup (both disordered and regular frequencies) we fix $g=0.4
g_c$, while for uniformly distributed frequencies we choose $g= 0.4
g_t$. This choice is safely far away from the transition point:
the correlations among different oscillators that may
arise when the coupling $g$ approaches $g_c$ or $g_t$ are still negligible.
However, we have verified that our results hold also for different coupling
values (in particular, $g=0.8g_c$, not shown here) in the non-synchronized regime. 
Simulations are performed
starting from random initial phases and discarding a transient of
$T_0=N\cdot 10^3$ time units. Ensemble averages in the disordered cases are
typically performed over $\Omega=10^2 $ different frequency realizations,
while the single-sample LLE $\lambda_\alpha$ is computed over
$T = 2 N \cdot 10^3$ time units. We have verified that, with
this choice, the standard error of the single-sample LLE is at least
one order of magnitude smaller than the characteristic sample-to-sample 
spread of $\lambda_\alpha$. Therefore, we can confidently
express the numerical uncertainty of our estimate of $\lambda (N)$
with the sample-to-sample standard error,
\begin{equation}
S_E = \frac{\Delta \tilde\lambda}{\sqrt{\Omega-1}} \; ,
\label{se}
\end{equation}
where $\Delta \tilde\lambda$ is the standard deviation of the $\lambda_{\alpha}$s.

In the regular setups, the uncertainty on $\lambda$ is only due to the temporal fluctuations 
of the FTLE $\lambda_t$. In these cases we define the uncertainty as
\begin{equation}
s_E = \frac{\Delta \lambda}{\sqrt{n-1}} \; ,
\label{se2}
\end{equation}
where $\Delta \lambda$ is the standard deviation of $\lambda_\tau$, $\tau$ is a long enough time
to ensure a statistical independence of two consecutive measurement, while $n=T/\tau$ is the number of 
data point available from a simulation of length $T$ (this latter time is typically chosen on the order of
$10^6 \sim 10^7$ time units).
In the remaining of this paper the error bars correspond to one standard error.

\begin{figure}
\includegraphics[scale=0.38]{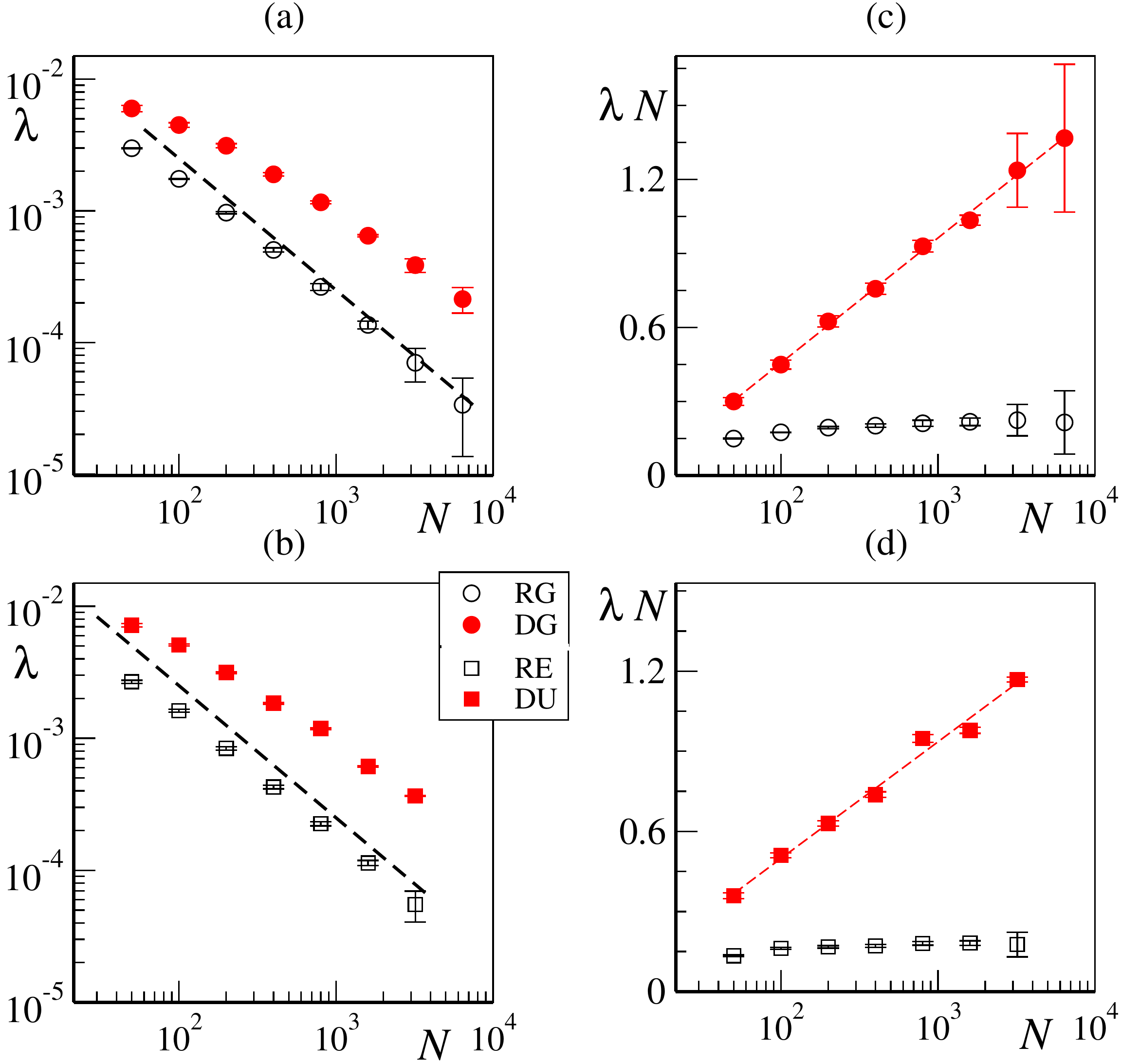}
\vspace{0.2cm}
\caption{Finite size scaling of the largest Lyapunov exponents in the incoherent phase (see text). (a) LLE $\lambda$
vs. system size for disordered Gaussian (DG, full red circles) and  regular Gaussian (RG, empty circles) frequencies.
(b) Same as (a), but for disordered uniform (DU, full red squares) and regular equispaced (RE, empty squares) frequencies.  
Axis are in double logarithmic scale, and the black dashed lines
marks a decay $\sim 1/N$. (c)-(d) The LLEs of panels (a)-(b) 
are multiplied by system size $N$ to better show the leading logarithmic
correction to the disordered frequencies choices. Axis are now in log-lin scale, 
and the red dashed lines represents the best logarithmic fit of the disordered frequencies data.}
\label{fig1}
\end{figure}

The numerical results are resumed in Fig.~\ref{fig1}a-b, where the scaling of the LLE with $N$ is
shown for the above-mentioned setups. A common property of all simulations is that $\lambda$ decreases
with the system size. This is not surprising since, below threshold, in the thermodynamic limit,
the oscillators are mutually uncoupled.
A less trivial result is that the regular setups (open black symbols) yield a substantially smaller LLE
than the disordered ones (full red symbols). The difference is not only quantitative, but even qualitative:
a look at Fig.~\ref{fig1}c-d reveals that while the LLE decreases as $1/N$ in the regular cases,
it behaves as
\begin{equation}
\lambda (N)  \sim \frac{\ln N}{N} \,,
\label{log-scaling2}
\end{equation}    
in both disordered cases.

It is rather difficult to push the direct numerical simulations of the full 
model (\ref{Kuramoto2},\ref{KTS2}) beyond $N=10^4$. In fact, the increasing demand of CPU time
to simulate larger ensembles is accompanied by the need to ensure a constant relative accuracy
of the corresponding decreasing LLEs.

In regular setups the problem is worsened by the additional slower decay of the autocorrelation function.
In fact, the statistical error, estimated from Eq.~(\ref{se2}) increases, due to the increase of
the autocorrelation time, which reduces the number $n$ of statistically independent points,
within a given time range $T$.
The problem is even more pronounced for equispaced frequencies (\ref{eq::eqspaced}), since in 
the thermodynamic limit $R=0$ and the dynamics is periodic with a period $T_P=2\pi\,N$,
as it easily seen from Eq.~(\ref{eq::eqspaced}).
Finite system sizes (and, thereby, non-zero coupling strengths), of course, destroy the exact periodicity, 
but we have observed that correlations between consecutive periods persist 
(up to around 5 periods for our coupling choice) and should be accounted for when estimating $n$. 
The same effect is also present for regular Gaussian frequencies, although
the periodicity is, in this case, approximate even in the zero-coupling case. 
In fact, upon increasing $N$, the frequency distribution is increasingly uniform on microscopic scales.
In the central, densest region, expanding the error function in Eq.~(\ref{eq::detdistrib2}) up to first
order in the argument, one finds $T_P \approx 2\sqrt{2\pi}\,N$.

\begin{figure}
\includegraphics[scale=0.37]{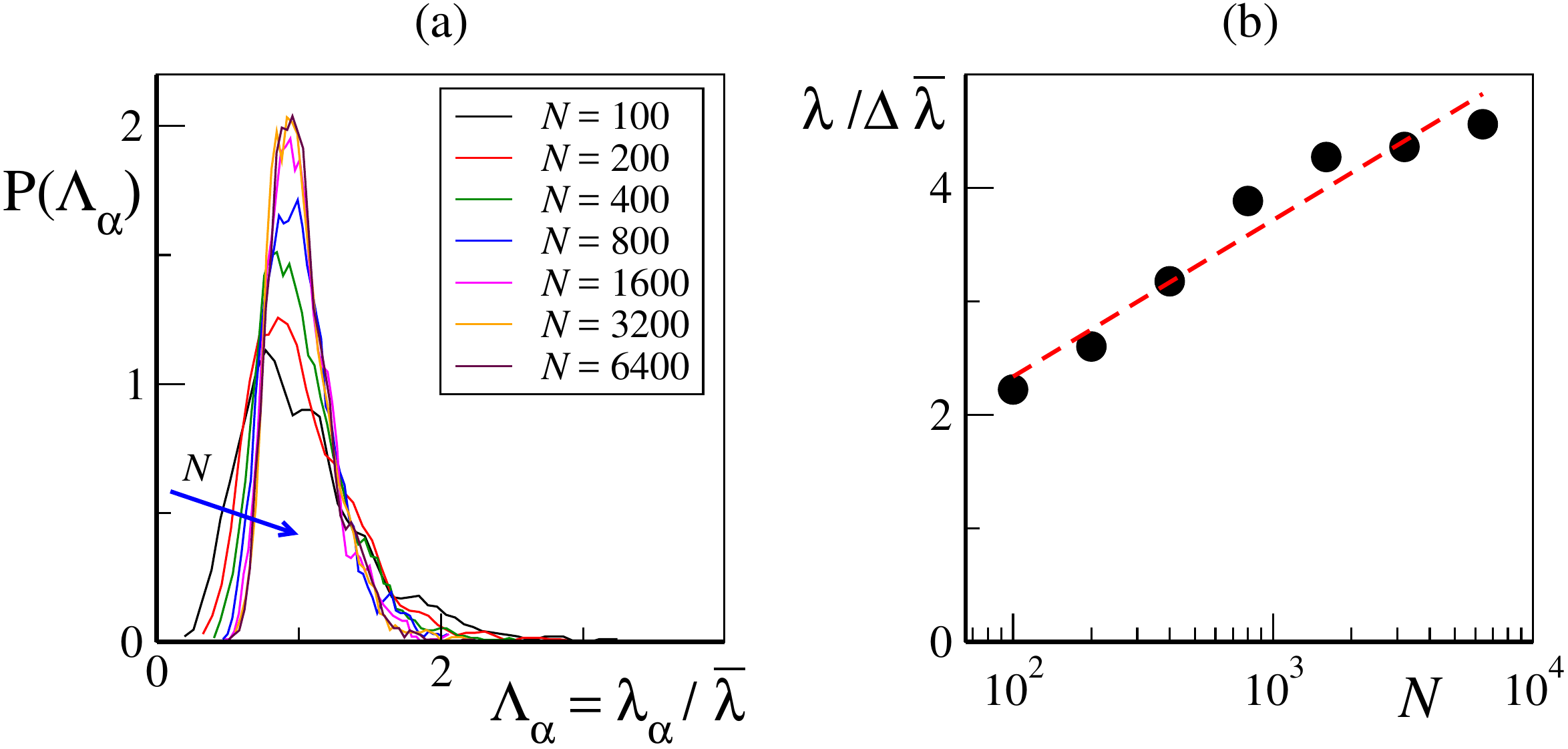}
\caption{Marginal self averaging properties. (a) Probability distributions for the
rescaled LLE $\Lambda_\alpha$ (see main text) for different system sizes (size is increasing
along the blue arrow). Each probability distribution is estimated from $10^3$ realizations of the disordered Gaussian frequencies. (b) Inverse ratio $\lambda / \Delta \bar{\lambda}$ vs. system size in a log-lin scale. 
The dashed red line marks a logarithmic fit.}
\label{fig2}
\end{figure}

In disordered setups, these coherence problems are not so crucial, but it is necessary to deal with
sample-to-sample fluctuations. Actually, this issue has a conceptual relevance: so long as the
relative amplitude of these fluctuations decreases with $N$, one could conclude that the LLE is a well defined,
self-averaging quantity; otherwise, one should conclude that sample-to-sample fluctuations remain
relevant in the thermodynamic limit.
We have explored this issue, by determining the distribution of LLEs for different
values of $N$. The results are illustrated in Fig.~\ref{fig2}a, where $\lambda_\alpha$ is rescaled
to the average value ($\Lambda_\alpha \equiv \lambda_\alpha/\lambda$), so as to compare the size of
the fluctuations for different system sizes. The distribution appears to narrow, but the dependence on
$N$ is very slow. A more quantitative analysis can be performed by
plotting the inverse of the relative standard
deviation $\Delta \bar{\lambda}/\lambda$ versus $N$. The data plotted in Fig.~\ref{fig2}b suggest
a logarithmic convergence to zero, i.e. a marginally self-averaging property, which
makes large-size accurate simulations rather problematic.

Altogether, we have found numerical evidence that the LLE in the incoherent 
phase of the Kuramoto model decays to zero as $1/N$. This result agrees with earlier numerical estimates
provided in Ref.~\cite{Popovych} for regular equispaced frequencies. On the other hand, our simulations suggest
that disordered setups (like the original Kuramoto model) are characterized by a logarithmic correction.
As the conjectured existence of two universality classes is only based on numerical simulations,
it is desirable to consider as large systems as possible to avoid being misled by uncontrollable
finite-size effects.
In order to reach larger $N$ values it is necessary to alleviate the simulation burden. 
This is precisely the goal of next section.

\section{Approximate models}

In order to gain insight on the scaling behavior of the LLE discussed in the previous section,
here we introduce some approximations/variants of the original model.

\subsection{Quasiperiodic phase-space approximation}

In the thermodynamic limit, i.e. for a strictly infinite $N$, the oscillators are uncoupled
and the overall dynamics is quasiperiodic (QP). For a large but finite $N$, the order parameter
is small but nonzero: this induces fluctuations around the QP dynamics and modifies
the tangent-space dynamics.
In order to test to what extent the two effects contribute to the scaling behavior of the LLE 
we have decided to investigate a setup where the the phase-space dynamics is purely QP,
\begin{equation}
\dot{\theta}_i = \omega_i \, .
\label{qp}
\end{equation}
Before proceeding in that direction, it instructive to quantify the effect of the coupling,
computing the frequency shift $\Delta \omega_j \equiv \langle \dot \theta_j(t) \rangle_t - \omega_j$ 
(here $\langle \cdot \rangle_t$ denotes time average).

\begin{figure}[t!]
\includegraphics[scale=0.36]{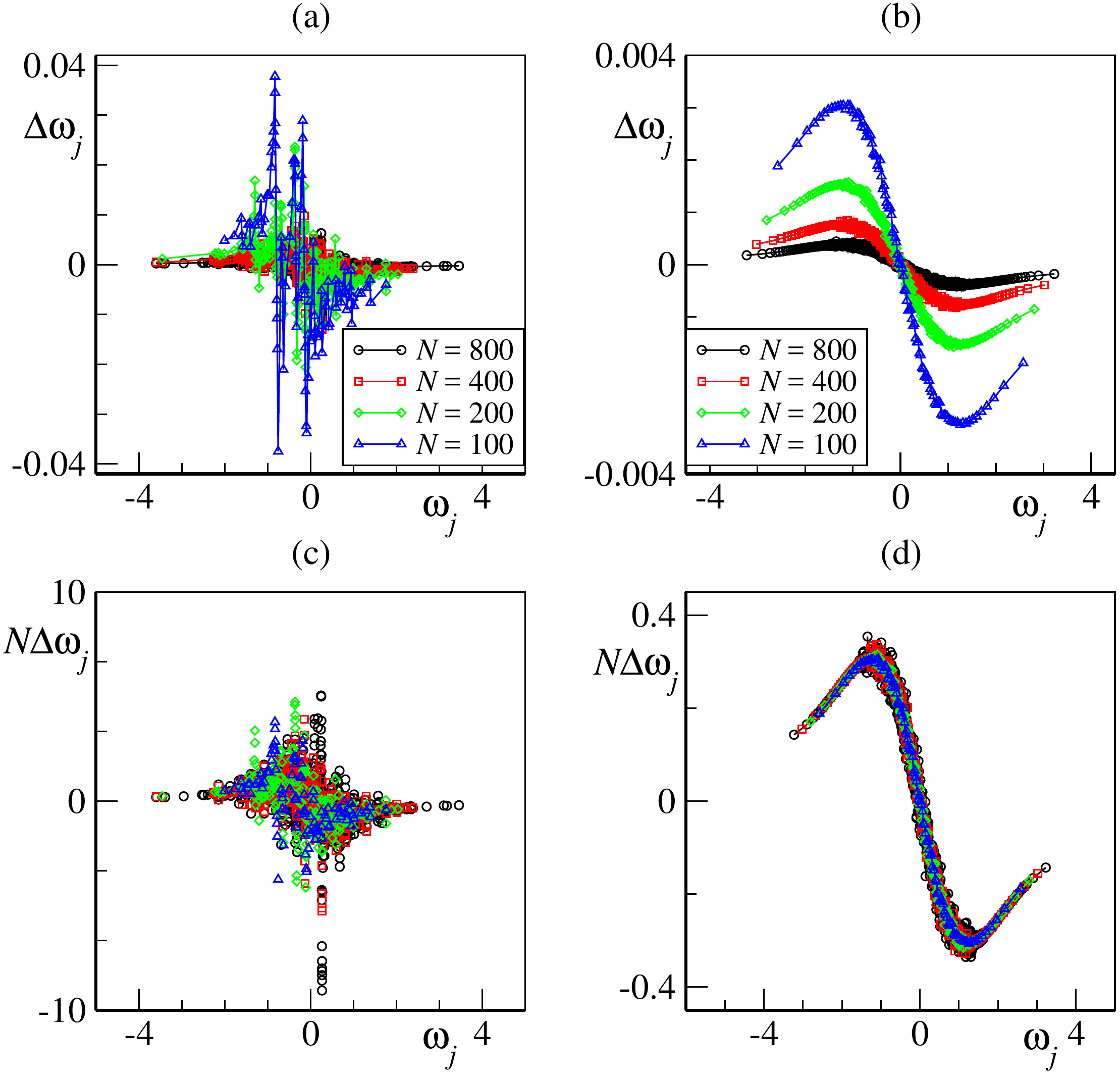}
\vspace{0.2cm}
\caption{(a) Frequency shift $\Delta \omega_j$ vs. the natural
  frequency $\omega_j$ for different system sizes $N$
for one typical realization of DG frequencies and an average time of $T
\approx 10^6$ time units. (b) same as (a), but for RG frequencies. (c)-(d) Frequency shift
multiplied by the system size $N$ to show the finite size scaling $1/N$.}
\label{fig3A}
\end{figure}

In Fig.~\ref{fig3A} we plot $\Delta \omega_j$ versus the natural frequency for the two Gaussian setups 
(DG, panel a; RG panel b). DG shows larger deviations in the central (denser) part of the frequency range. 
This actually reflects the tendency of the oscillators to form partially synchronized
clusters, where the frequency spacing $\Gamma_j\equiv\omega_{j+1}-\omega_j$ tends to be smaller (a precursor of
the synchronization transition).
The RG setup behaves differently: the ``response'' curve is symmetric (an obvious consequence of the 
perfect symmetry of the RG distribution) and the frequency shift much smaller, by roughly one
order of magnitude. From the overlap in the lower panels, we can appreciate the scaling with $N$ of the
frequency shift. The numerics suggests that $\Delta \omega_j \sim 1/N$ in both cases, 
although the scaling is much cleaner for RG. 
A possible motivation for the quantitative differences exhibited by the two setups comes from the
distribution of the frequency spacings $\Gamma_j$. Because of the quenched randomness, in the DG case, 
$\Gamma_j$ can be occasionally very small; in such a case the $j$th and $j+1$st oscillators 
act almost as a single unit within an ensemble of $N/2$.

\begin{figure}[t!]
\includegraphics[scale=0.36]{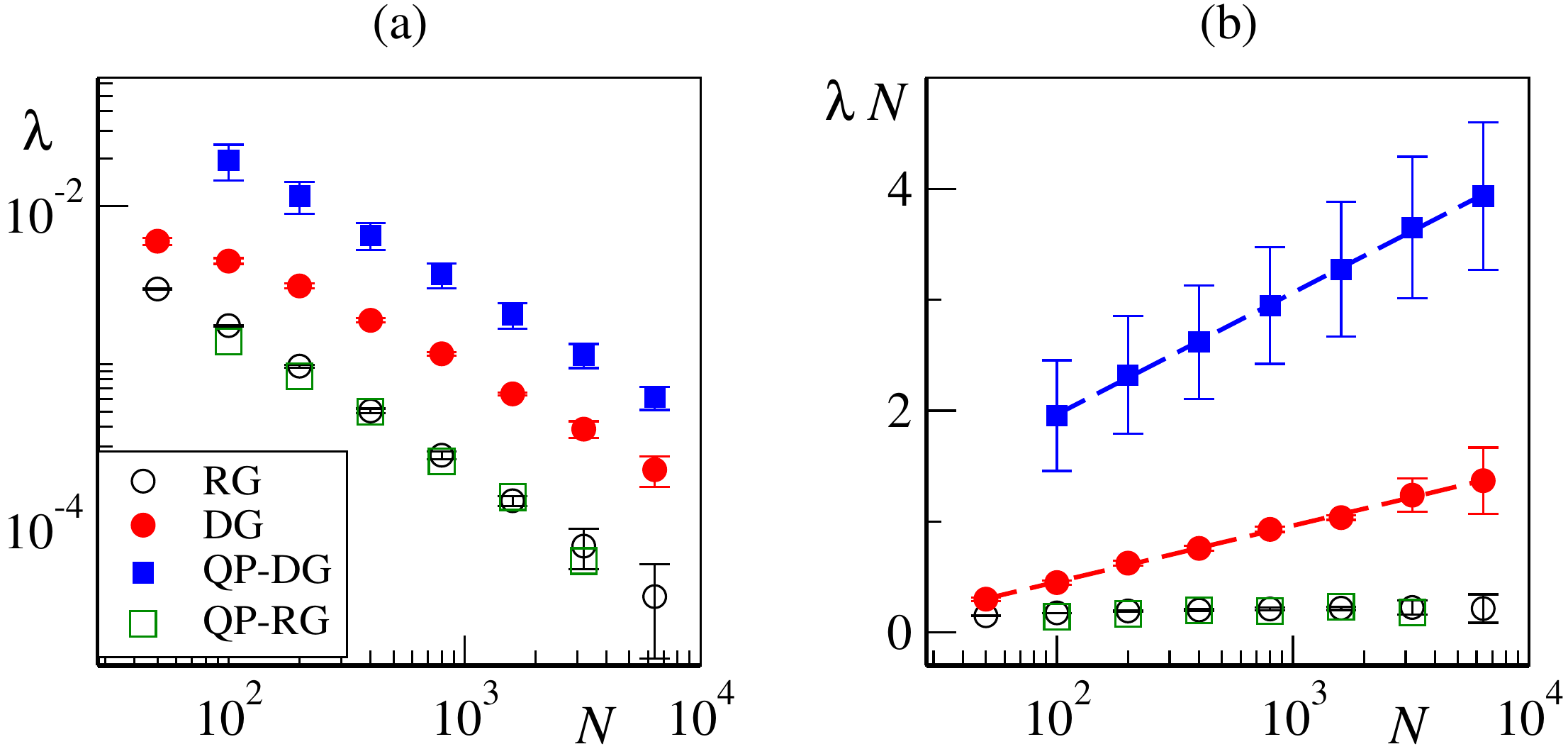}
\vspace{0.2cm}
\caption{Quasi periodic approximation (a) LLEs vs. system size for
the quasi-periodic (squares) and full dynamics (circles) in the
disordered (full symbols, data averaged over 100 different frequency realizations) and regular (open symbols)
Gaussian frequencies set-ups. 
Axes are in a doubly logarithmic scale. (b) The LLE of panel (a) are multiplied 
by system size $N$ to highlight the leading logarithmic correction to the disordered frequencies set-ups. 
Axes are now in log-lin scale, and the red dashed lines represents 
the best logarithmic fit of the disordered frequencies data.}
\label{fig3B}
\end{figure}

The above numerical studies indicate that deviations from the QP zero-coupling behavior are of
order $1/N$. Under the assumption that such deviations do not crucially affect the scaling
behavior of the LLE, we focus on the tangent-space evolution ruled by Eq.~(\ref{KTS2}),
feeding it with the QP evolution of the single phases.
In fact, the results shown in Fig.~\ref{fig3B}a-b, confirm that the regular setup is
characterized by a $1/N$ scaling, while the disordered one, exhibits a $(\ln N)/N$ behavior.
Additionally, in the disordered case, the LLE is much larger (by about one order of magnitude)
in the QP case, with an obvious numerical advantage. 

\subsection{Discrete-time approximation}

\begin{figure}[t!]
\includegraphics[scale=0.45]{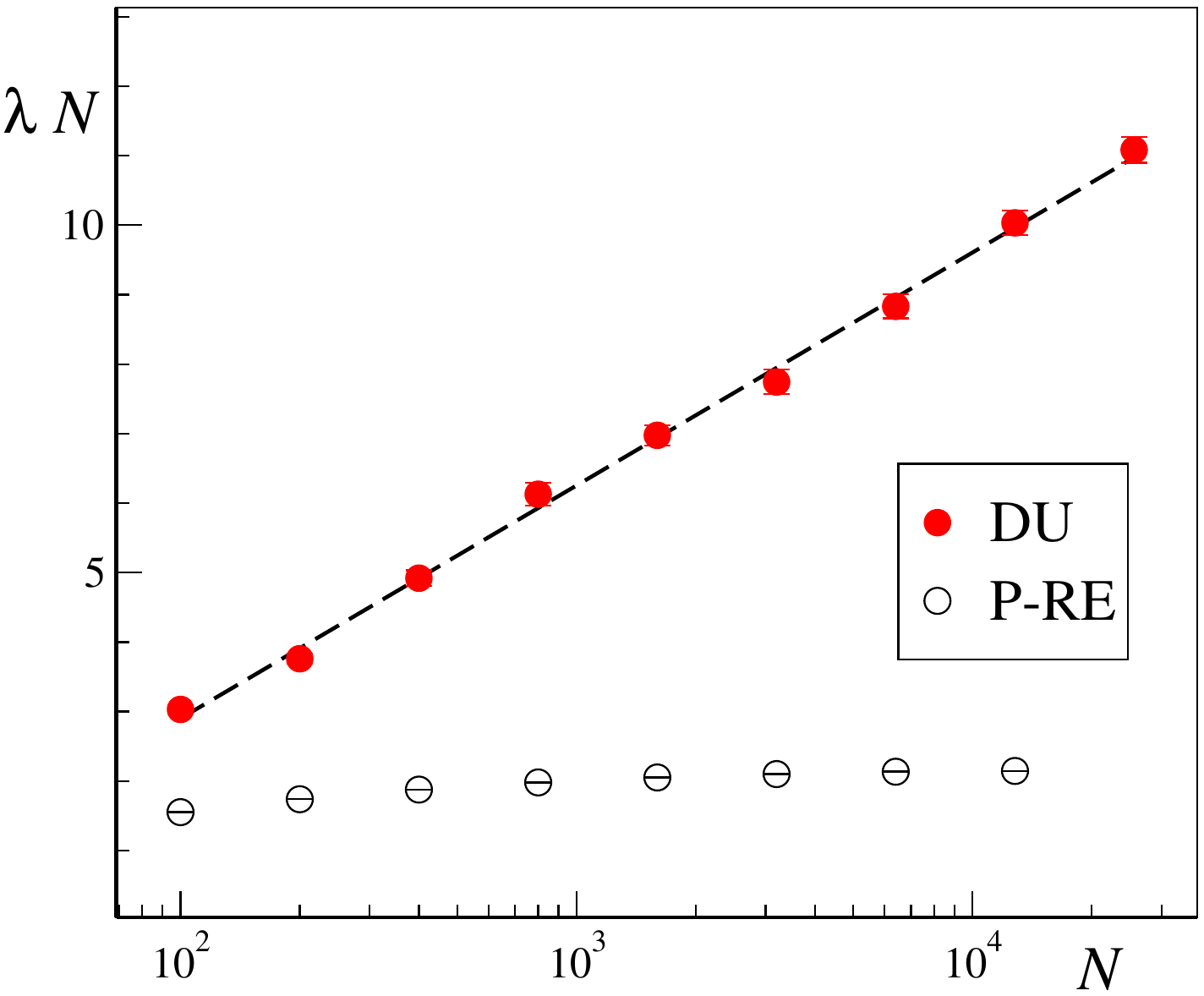}
\vspace{0.15cm}
\caption{Rescaled largest Lyapunov exponent for the
  discrete Kuramoto model vs. system size. Disordered uniformly
  distributed (DU) frequencies are marked by full red circles, while
  empty black squares refer to perturbed ($p=1$) regular equispaced 
  frequencies. Axes are
in lin-log scale to highlight the logarithmic correction to the
disordered ensemble (black dashed line fit). Data has been averaged
over 100 different realizations.}
\label{fig4}
\end{figure}

While the quasiperiodic approximation has the advantage of letting us dealing with larger LLEs, 
thus reducing the statistical fluctuations, it has to be reckoned that it does not
speed up significatively the numerics, as it does not free us from the burden 
of integrating continuous-time differential equations. 
Under the conjecture that the logarithmic correction is a universal property
of a large class of disordered, coupled phase-oscillators, we have further simplified
the model. In practice, we have considered discrete-time setups, which are 
significantly faster to simulate (by at least two orders of magnitude) 
and thereby allow for more accurate numerical tests. The tangent-space evolution
is obtained by discretizing the Kuramoto model, again under the assumption of
a quasi-periodic dynamics in real space, $\theta_i = \alpha_i + \omega_i t $,
\begin{eqnarray}
\delta \theta_i(t+1) &=& [1- g R(t) \cos(\psi-\omega_it-\alpha_i)] u_i + \nonumber \\
&&gZ(t) \cos (\beta- \omega_i t-\alpha_i) \; .
\label{eq:kura_disc}
\end{eqnarray}
By assuming a disordered uniform distribution of frequencies in the interval $[-1/2,1/2]$ 
and a uniform distribution of the initial phases $\alpha_i$, 
together with setting $g=1/2$ (we have separately verified that the dynamics stays incoherent 
for this coupling), we have been able to extend the numerical evidence of the class II
behavior to much larger system sizes, as it can be appreciated 
in Fig.~\ref{fig4} (red dots). We are therefore more confident in conjecturing 
that the logarithmic correction survives in the thermodynamic limit.

Given the opportunity to perform large-size simulations offered by the discrete-time
setup, we have decided to explore how large is the universality class characterized
by the presence of logarithmic corrections.
We have already seen that in regular distributions the LLE scales simply as $1/N$. 
In the case of uniform distributions, it is not advisable to investigate equispaced 
frequencies in the QP approximation, as this would result in a strictly 
periodic phase-space dynamics, that is obviously non generic.

This pathological property has suggested us to consider a variant of the regular
distribution, where a quenched microscopic randomness is included,
\begin{equation}
\omega_j = \frac{2j-1 + p \,\zeta_j}{2N} - \frac{1}{2} 
\label{eq::eqspaced2}
\end{equation}
where the $\zeta_j$'s are independent and identically distributed (iid) random variables distributed in $[-1,1]$,
while $p$ quantifies the srength of the perturbation.
In the following we refer to this setup as Perturbed Regular Equispaced (P-RE).
For $p=0$, the distribution is perfectly regular. For $0<p<1$ the 
frequency spacing $\Gamma_j$ is strictly larger than $(1-p)/N$;
this can be interpreted as the minimal separation between
consecutive frequencies. For $p=1$ such a gap vanishes.

Numerical simulations of the P-RE setup (\ref{eq:kura_disc}) 
reveal that the ensemble averaged LLE $\lambda$ scales as $1/N$ 
(see the black squares in Fig.~\ref{fig4} for the case $p=1$).
Therefore, we can conclude that the presence of stochastic fluctuations
is not a sufficient condition to observe the logarithmic correction.

It should, however, be noticed that the P-RE distribution differs
from the DU one, in the same way as the configuration of a one-dimensional
solid differs from a one-dimensional gas (interpreting the frequency values
as the positions of ``atoms" on a line): only the former is characterized
by long-range order. This leads to conjecture that the
absence of long-range order might be a key property.

\subsection{Frequency spacing distributions}
\label{FSD}

So far, we can conjecture that long-range order in the frequency
distribution suppresses the logarithmic correction to the scaling of the
LLE. Can we conclude, that the absence of long-range order is a sufficient condition for 
the scaling $\lambda \sim \ln N/N$?

In order to shed light on this issue, we now introduce and analyze further classes
of frequencies ensembles. As a matter of fact, instead of generating 
directly the $N$ frequencies, we now choose to generate directly the $N-1$
frequency spacings $\Gamma_j=\omega_{j+1} - \omega_j$ from a given
probabilty distribution $\Pi (\Gamma)$. The frequencies are afterwards shifted and scaled
in order to fill the $[-1/2, 1/2]$ interval. 

The selected distributions are graphically depicted in Fig.~\ref{fig5}a, while
their mathematical expression can be found in the appendix~\ref{app}.
They have been chosen with the goal of clarifying the role of quasi-degeneracies,
by varying the density close to zero.
So, we go from (i) the inverse square-root, characterized by a divergence in zero, to (ii) the
Poisson (which, incidentally, corresponds to the original DU), and (iii) flat distributions,
characterized by a finite density in zero, and to (iv) the linearly vanishing density of
the triangular and pyramidal distributions. Finally, we have also included
a modified pyramidal density with a minimum frequency spacing $b$.
It should be noticed that by construction no long-range order is ever present.

\begin{figure}[t!]
\resizebox{211pt}{!}{\includegraphics{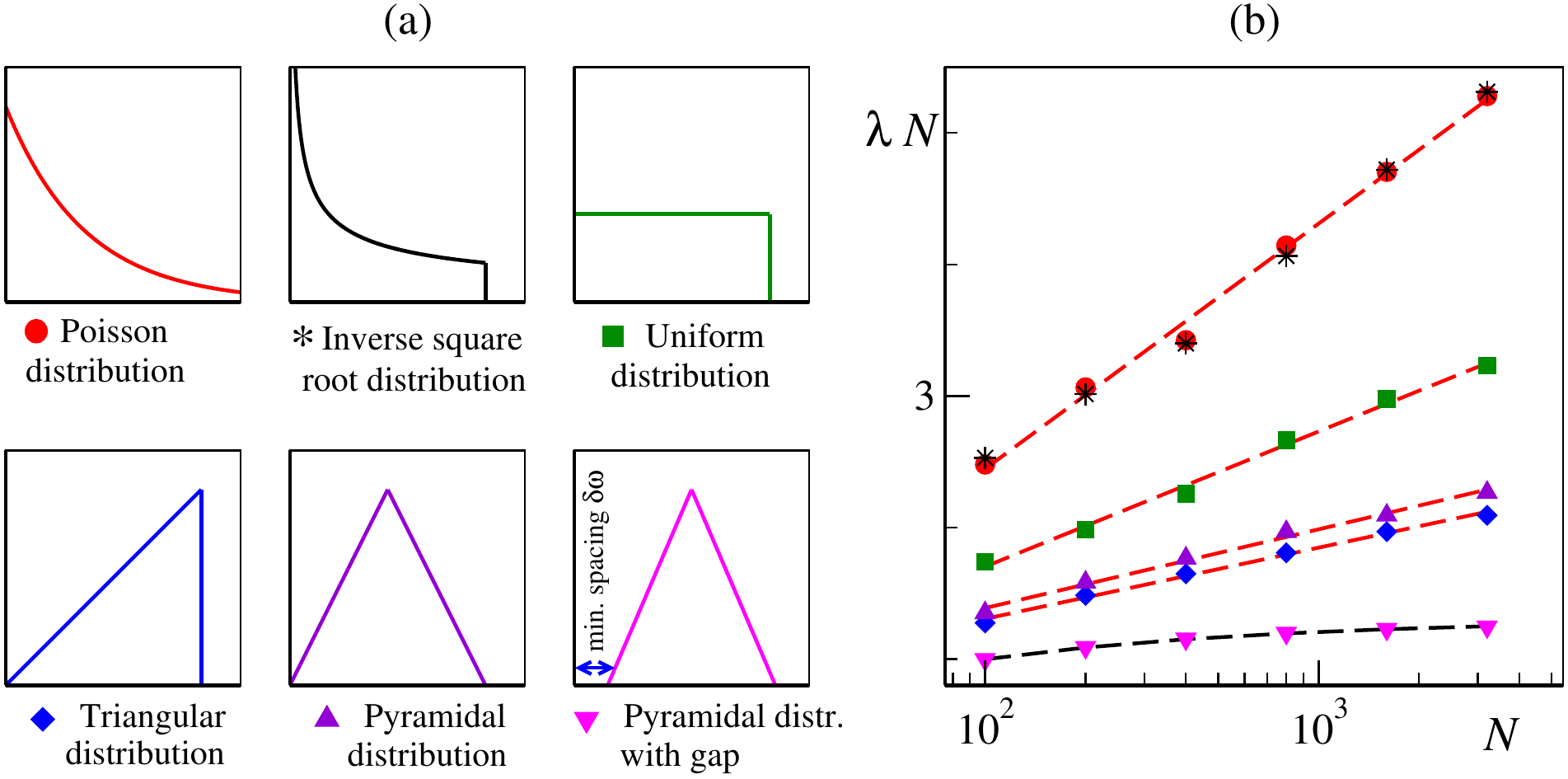}}
\vspace{0.2cm}
\caption{Quasi-periodic Kuramoto model with disordered frequencies
  ensembles generated via the frequency spacing distribution method (see
  text).
(a) Graphical depiction of the different frequency spacing
  distribution tested here. (b) Rescaled largest Lyapunov
  exponent. The dashed lines have been obtained as the best fit of the
numerical data (see text). Data has been averaged
over 100 different realizations.}
\label{fig5}
\end{figure}

Our results, reported in Fig.~\ref{fig5}b for the quasi-periodic
dynamics and tangent space coupling $g=0.4$, indicate that upon decreasing the amount of
quasi-degeneracies (i.e. the spacing probability $\Pi$ near $\Gamma=0$), 
the LLE tends to become smaller, confirming the heuristic idea
that they contribute to increasing the degree of instability.
However the overall picture is not as simple as one would like: in all cases
the logarithmic correction is present (the best fits for $\lambda N$ are
logarithmic in $N$), with the exception of the gapped pyramidal
case, where our best fit gives $\lambda N = \Lambda_0 + b/\sqrt{N}$.
In other words, only when an identically zero-density appears, the quantitative
decrease of the LLE transforms into a qualitative change of the scaling behavior.
Therefore, we see that the lack of long-range order alone is not sufficient to
induce the logarithmic correction: The presence of finite gaps in the
frequency spacing seems to suppress the logarithmic correction even in
the absence of long-range order.

\section{Theoretical considerations}
\label{theo}

In this section we approach the problem from an analytical point of view.
We concentrate on the tangent-space dynamics described by Eq.~(\ref{KTS2}), 
noting that it is determined by two contributions: a multiplicative
term $-gR\cos(\psi-\theta_i) \delta \theta_i $ and a common pseudo-additive term 
$g Z \cos (\beta- \theta_i)$ which couples the different tangent-space
variables through their wheighted average (\ref{eq:wa}). Note that
also this latter term is linear in ${\bm \delta \theta}(t) \equiv \left(
\delta \theta_1, \delta \theta_2, \dots, \delta \theta_N \right)$, so
that Eq.~(\ref{KTS2}) is homogeneous in ${\bm \delta \theta}(t)$, as it is expected.
 
\subsection{Full stochastic approximation}

In the incoherent phase, under the assumption of a stochastic dynamics of both the Kuramoto
order parameter $Re^{i \psi}$ and the weighted tangent-vector average $Z\mathrm{e}^{i\beta}$, 
it is possible to derive analytical results,
which do not only provide an approximate description of the full model, but help also to
appreciate the role of the quenched distribution of frequencies in the tangent space evolution.

We start replacing $-g R \cos(\psi-\theta_i)$ and $g Z \cos (\beta- \theta_i)$ in
Eq.~(\ref{KTS2}) with two stochastic terms, 
\begin{equation}
\dot{u}_i = \xi_i(t)u_i + \sqrt{\langle u^2 \rangle} \eta_i(t) \; ,
\label{eq:stoch1}
\end{equation}
where $\xi_i$ and $\eta_i$ are independent white noises characterized by variances
$\sigma^2_\xi$ and $\sigma^2_\eta$, respectively. The square root term is factored out
to show explicitly that the above equation is homogeneous. (For
convenience, in the stochastic approximation we denote the tangent
vector components $\delta \theta_i$ of the original problem as
$u_i$.)

By further assuming the Stratonovich interpretation for 
the multiplicative noise, the corresponding Fokker-Planck equation for
the probability distribution $\rho(u, t)$ \cite{FP} is
\begin{equation}
\frac{\partial \rho}{\partial t} = -\frac{\partial}{\partial u}
\left[ \frac{\sigma_\xi^2}{2}u\rho - \frac{\partial}{\partial u}
\left( \frac{\sigma_\xi^2}{2}u^2 + \frac{\sigma\eta^2}{2}\langle u^2\rangle
\right)\rho
\right]\; .
\label{eq:fp}
\end{equation}
At this level, the presence of a positive Lyapunov exponent $\lambda$ is signalled by an exponential broadening of the distribution 
$\rho(u,t)$.
If we introduce the rescaled variable
\begin{equation}
v = k u \mathrm{e}^{-\lambda t} \; ,
\label{eq:change}
\end{equation}
the probability density $S(v)dv = \rho(u,t) du$ is independent of time. Simple calculations show that
\begin{equation}
\frac{\partial}{\partial v}
\left[ \lambda v S - \frac{\sigma_\xi^2}{2}v S + \frac{\partial}{\partial v}
\left( \frac{\sigma_\xi^2}{2}v^2 + \frac{\sigma_\eta^2}{2}
\right)S
\right] =0 \; ,
\label{eq:FP_stat}
\end{equation}
where we have chosen the normalization constant $k$ so that $\langle v^2 \rangle = 1$.
By imposing the usual no-flux condition, the above equation reduces to
\begin{equation}
\left (\lambda + \frac{\sigma_\xi^2}{2}\right )v S +
\left( \frac{\sigma_\xi^2}{2}v^2 + \frac{\sigma_\eta^2}{2}\right )
\frac{\partial S}{\partial v} = 0 \; ,
\label{eq:FP2_stat}
\end{equation}
which can be solved, obtaining a sort of generalized Lorentzian shape,
\begin{equation}
S(v) = \frac{C}{(a^2+v^2)^{\tilde{\lambda}+1/2}} \; .
\label{eq:FP_sol}
\end{equation}
where 
\begin{equation}
a = \sigma_\eta/\sigma_\xi   \qquad  \overline \lambda = \lambda/\sigma^2_\xi \; .
\label{params}
\end{equation}
We are left with two undetermined constants, $C$ and
$\lambda$. However, they have to satisfy the conditions 
\begin{equation}
\int_{-\infty}^{\infty} S(v) dv = 1 \;\;\;\;\mbox{and}\;\;\;\;\
\langle v^2\rangle =\int_{-\infty}^{\infty} v^2 S(v) dv = 1\,.
\label{eq:norms}
\end{equation}

Upon normalizing the area to unity, for finite $a$ one finds that
\begin{equation}
S(v) = \frac{a^{2\overline \lambda}}{B(1/2,\overline \lambda)(a^2+v^2)^{\overline \lambda+1/2}} \; ,
\label{eq:FP_sol2}
\end{equation}
where $B(x,y)$ is the Beta function.
Finally, $\overline \lambda$ is determined by imposing the
self-consistency condition ${\langle v^2\rangle \!=\! 1}$, 
\begin{equation}
\frac{B(3/2,\overline \lambda-1)}{B(1/2,\overline \lambda)}a^2=1 
\label{eq:lambda}
\end{equation}
By making use of the connection between $B(x,y)$ and $\Gamma(x)$, the above equation reduces to
\[
\overline \lambda = 1 + \frac{a^2}{2}
\]
which, in the original framework, becomes
\begin{equation}
\lambda = \sigma_\xi^2 + \frac{\sigma_\eta^2}{2} \; .
\label{eq:lyap_theo}
\end{equation}
Note that this result also holds for $\sigma_\xi \to 0$, i.e. for a
vanishing contribution of the diagonal stochastic
term. In this case, in fact,  Eq.~(\ref{eq:FP2_stat}) admits the
Gaussian solution
\begin{equation}
S(v) = C e^{-\lambda v^2/\sigma_\eta^2} 
\label{eq:FP_solG}
\end{equation}
which, upon imposing the conditions (\ref{eq:norms}), gives
${\lambda=\sigma_\eta^2/2}$. 

The opposit limit of a vanishing noisy term $\eta$ is singular.
In fact, it is obvious that for $\sigma_\eta^2 \to 0$, in the absence of the second right hand side of
Eq.~(\ref{eq:stoch1}), the $u$ variables are mutually uncoupled and driven by a zero average term
$\xi$, so that the Lyapunov exponent must be zero. This singular behavior is indeed
signaled by the solution (\ref{eq:FP_sol}) being non-normalizable for
every value of $\lambda$.
We can thus conclude that, at least in our stochastic approximation,
chaos emerges from the fluctuations of the weighted tangent-vector average $Ze^{i\beta}$.

It is interesting to compare these results with the stochastic theory developed in
Ref.~\cite{GCpap} for globally coupled identical chaotic units. In such a case, the maximum 
Lyapunov exponent is strictly larger than the single-oscillator exponent 
by an amount equal to the half of the
variance of the single-oscillator LE \cite{NOTE-GC}. In the present setup, 
the fluctuations of the local Lyapunov exponent are represented
by $\sigma_\xi^2 $, while $\sigma_\eta^2$ simulates the fluctuating
coupling terms which needs to be present 
to ensure the increase of the LE. 

 \subsection{Discrete-time version}

\begin{figure}[t!]
\includegraphics[width=3. in,angle=0,clip=true]{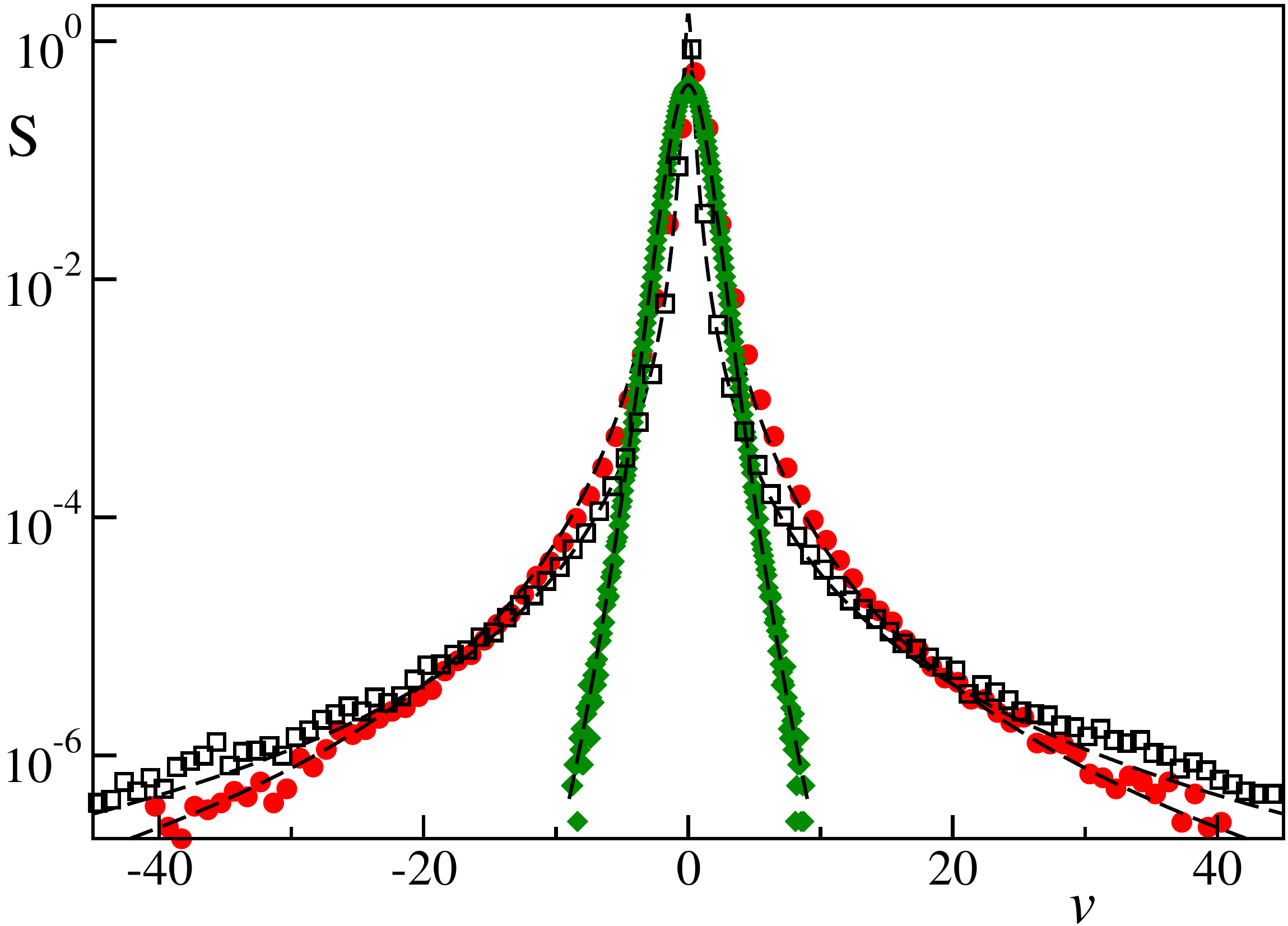}
\caption{Discrete time stochastic approximation. Numerical stationary
  probability distribution $S(v)$ (symbols) compared with the
  theory (dashed full lines). Simulations are performed by iterating a system of $4 \cdot 10^7$ elements with
$\sigma_\chi^2 = 10^2 f$ and $\sigma_\eta^2 = 10^2 (1-f)$, for $f$ such as
to have $a=3$ (green diamonds), $a=1$ (red circles) and $a=0.3$ (black
squares).}
\label{fig:distr}
\end{figure}

Next, we consider the discrete-time equivalent of Eq.~(\ref{eq:stoch1}),
\begin{equation}
u_i(t+1) = \exp(\chi_i(t)) u_i(t) + \sqrt{\langle u^2 \rangle} \eta_i(t)  \; ,
\label{eq:stoch_disc}
\end{equation}
where $\xi_i$ and $\eta_i$ are binary variables taking values 
$\pm \sigma_\xi$ and $\pm \sigma_\eta$ respectively.
This is basically the stochastic version of Eq.~(\ref{eq:kura_disc}), with the difference that
the term $1+\xi_i = 1- gR \cos(\psi-\theta_i)$ is replaced by $\exp(\chi_i)$.
In the limit $\chi,\xi\ll 1$ the two expressions are similar $(\langle \xi \rangle=0)$; 
however, in a discrete-time setup, the former one gives a non-zero contribution 
to the Lyapunov exponent, as it is seen by expanding the logarithm,
\begin{equation}
\delta \lambda = \langle \ln (1 + \xi)\rangle \approx -\frac{\langle \xi^2 \rangle}{2} \; .
\end{equation}
In order to avoid the presence of such a spurious contribution, we
prefer to adopt the exponential formulation. We perform numerical
simulations of (\ref{eq:stoch_disc}), comparing the 
stationary probability densities $S(v)$ with the Fokker-Planck
solution (\ref{eq:FP_sol2}). Parametrizing variances as
$\sigma_\chi^2 = 10^2 f$ and $\sigma_\eta^2 = 10^2 (1-f)$, from
Eqs. (\ref{params})-(\ref{eq:lambda}) (with the substitution $\xi \to \chi$)
we readily obtain $a^2=(1-f)/f$ and $\overline \lambda=(1+f)/(2f)$.

The simulations reported in Fig.~\ref{fig:distr} correspond to three
different values of the parameter $a$. The agreement confirms the
correctness of the theoretical analysis
in a discrete-time setup too.

\subsection{Finite-size scaling of fluctuations}

Now, we come closer to the Kuramoto-like setup. By comparing Eqs.~(\ref{KTS2}) and (\ref{eq:stoch1}),
we see that $\xi_i(t) = -R \cos(\psi-\theta_i)$, where $R$ is the usual Kuramoto order parameter.
In the asynchronous regime, $R$ is well known to be of order $1/\sqrt{N}$. 
By neglecting the correlations between $\psi$ and $\theta$, we expect $\sigma^2_\xi\approx 1/N$.
Analogously, one could also conjecture that $\eta_i(t) = \cos (\beta- \theta_i) R/\sqrt{\langle u^2 \rangle}$ should
be of order $1/\sqrt{N}$, resulting from the (weighted) average of $N$ terms of order 1. 
By further assuming a finite decorrelation time, from the theoretical formula (\ref{eq:lyap_theo})
one should then conclude that the Lyapunov exponent in a finite
incoherent system is expected to be of order $1/N$.

On the other hand, the simulations reported in the previous sections show that this is not the case
for disordered frequency setups, which clearly show a logarithmic correction to the scaling. 
While, there is no reason to challenge the claim that $\langle R^2(t)\rangle \approx 1/N$, 
as this is the signature of asynchrony in a finite system, 
we have monitored $Q_Z^2 \equiv \langle Z^2\rangle$ and 
$Q_\beta^2 \equiv \langle Z^2 \cos^2 (\beta- \theta_i)\rangle$ in the
discrete Kuramoto model (\ref{eq:kura_disc})
for different values of $N$ and different realizations of the
disordered, uniformly distributed (DU) bare frequencies.

\begin{figure}[t!]
\includegraphics[width=3.5 in,angle=0,clip=true]{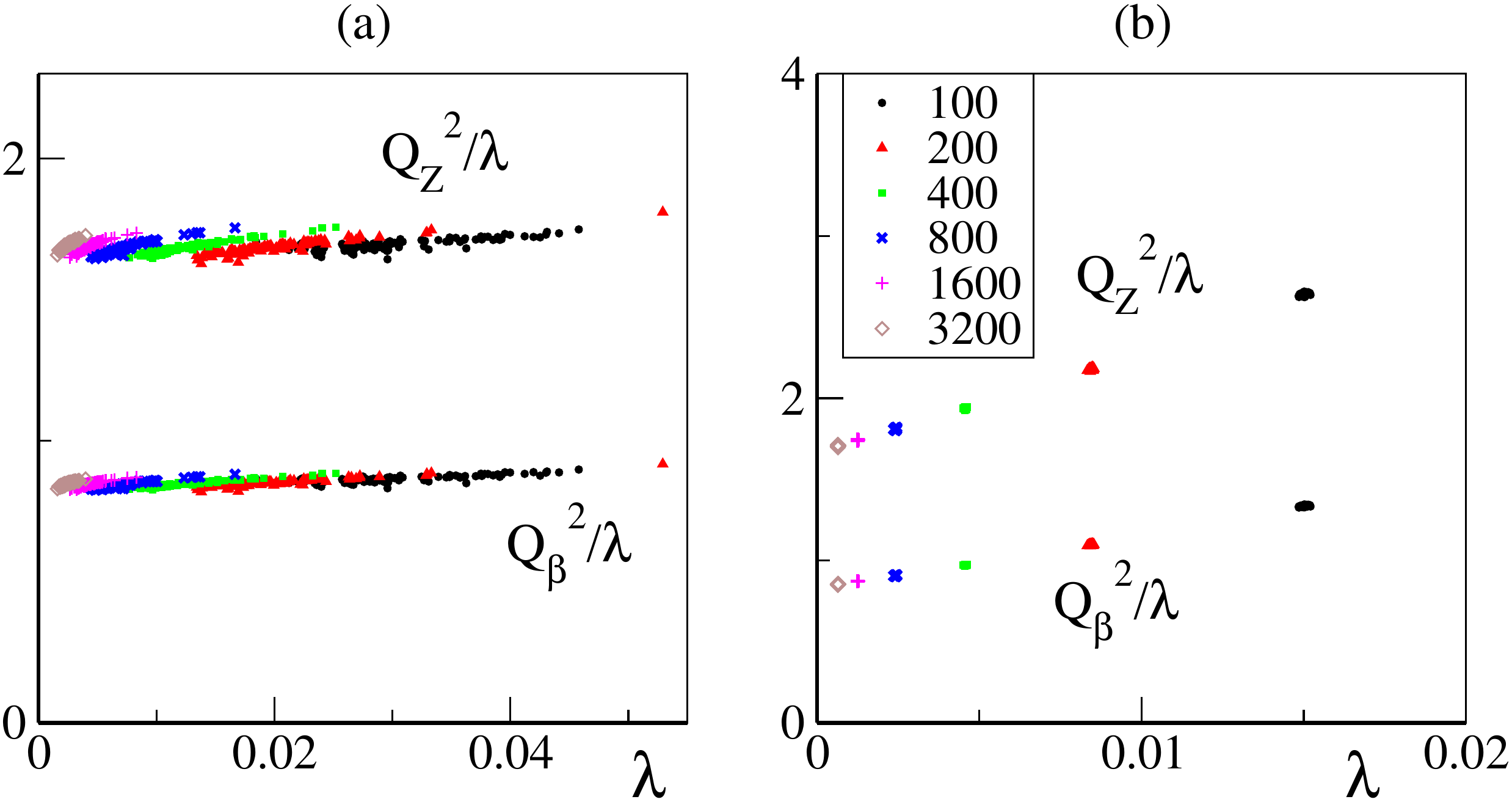}
\caption{The two average observables $Q_Z^2/\lambda$ and
  $Q_\beta/^2\lambda$ for different values of $N$ and 100 different realizations of
the frequencies in the discrete Kuramoto model
(\ref{eq:kura_disc}). (a) Disordered uniform frequency setup.  (b)
Perturbed regular equispaced frequencies ($p=0.8$). Single sample
time averages are performed over about $4 \cdot 10^5$ time units.}
\label{fig:cou_lambda}
\end{figure}

The results are presented in Fig.~\ref{fig:cou_lambda}a, where both $Q_Z^2$ and $Q_\beta^2$ are plotted versus
the single realization LLE $\lambda_{\alpha}$ (after being
divided by $\lambda_{\alpha}$ themselves). 
We notice a strong correlation between $Q_Z^2$ and
$Q_\beta^2$ with $\lambda_{\alpha}$, confirming our theoretical insight that
they are the basic sources for the scaling of the largest Lyapunov
exponent. Note also that no significative correlation is found between single-realization
 LLEs and the mean squared Kuramoto order parameter
$\langle R^2\rangle$ (not shown).
More unexpected is that the correlation persists across different values of $N$. 
The most important point, however, is that the two quasi-curves 
extrapolate to a finite value for $\lambda_{\alpha}=0$: this means that in
the thermodynamic limit (when $\lambda$ vanishes) $Q_R^2$ and $Q_\beta^2$ 
are proportional to $\lambda_{\alpha}$ itself:
in other words, as $\lambda \approx \ln N/N$, so are the two $Q^2$ variables, which are therefore
responsible for the anomalous scaling. Finally, note also that the ratio between the two quantities
is close to 1/2, revealing that the orientation of $Z$ is uncorrelated
with the local angle in real space.

For comparison, we also measured $Q_Z^2$ and $Q_\beta^2$ for a
discrete Kuramoto frequency setup that shows a clear $1/N$ scaling, namely the P-RE setup. 
Our results, reported in Fig.~\ref{fig:cou_lambda}b,
again show a strong correlation of $Q_Z^2$ and $Q_\beta^2$ with the
single realization LLE. However, due to the much smaller sample-to-sample fluctuations, 
in this case the data points for different system sizes are well separated, with a
limited variability of $\lambda_{\alpha}$ between
realizations. This result confirms that the fluctuations of 
$Z \mathrm{e}^{i\beta}$ are the main source of chaoticity also in this
different setup. However, the finite extrapolation
of the ratios $Q_Z^2/\lambda_{\alpha}$ and $Q_\beta/^2\lambda^{\alpha)}$ now indicates
that in this case the two $Q^2$ variables scale as $1/N$. 

To summarize, we have been able to trace the scaling of the LLE to the
behavior of the fluctuations of the weighted tangent vector average 
$Z \mathrm{e}^{i \beta}$. In typical disordered frequency setups one has 
\begin{equation}
\langle Z^2 \rangle = \left\langle \left | \frac{1}{N} \sum_j  \delta \theta_j
\mathrm{e}^{i\theta_j} \right|^2 \right\rangle \sim \frac{\ln N}{N}
\label{eq:wa-scaling1}
\end{equation}
while for more regular distributions, characterized by long-range order and/or
non-zero minimum frequency spacings, one has 
\begin{equation}
\langle Z^2 \rangle \sim \frac{1}{N}
\label{eq:wa-scaling2}
\end{equation}

\subsection{Weak synchronization and localization in tangent space}

\begin{figure}[t!]
\includegraphics[width=3.5 in,angle=0,clip=true]{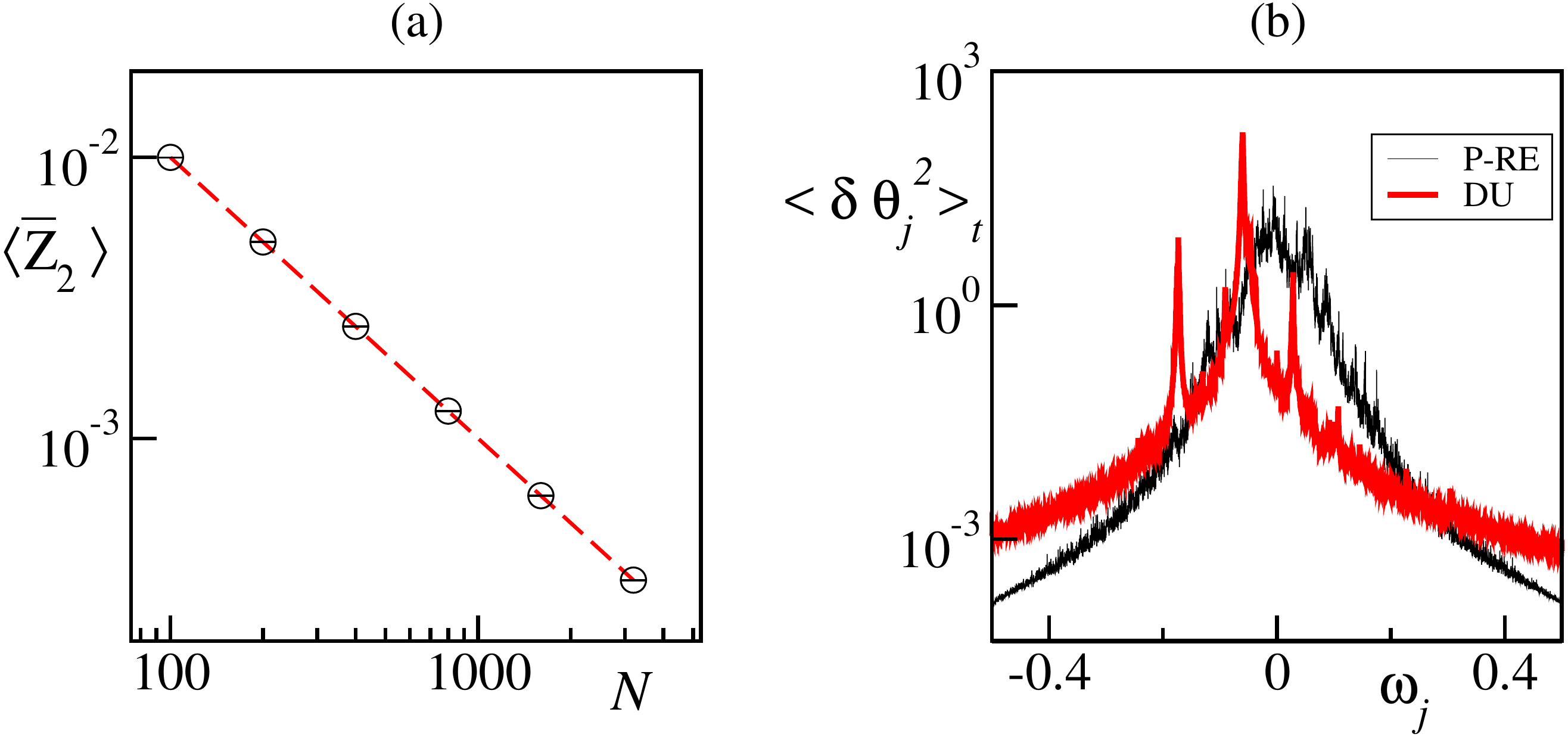}
\caption{(a) Finite size scaling of the reshuffled average
  $\tilde{Z}^2$ for DU frequencies (see text) in the discrete Kuramoto model.
Ensemble averages are taken over 100 different realizations of
the frequencies.
(b) Time-averaged (over $T=4 \cdot 10^5$ time units) 
square amplitude of the Lyapunov vector components as a function of
the corresponding natural frequency for a single frequency realization.}
\label{fig:proj}
\end{figure}

Finally, we discuss the physical interpretation of the deviations of
$\langle Z^2 \rangle$ from a $1/N$ scaling.

To illustrate the point, consider a disordered frequency setup where
$\langle Z^2 \rangle \sim \ln N/N$ and define
the ``reshuffled average''
\begin{equation}
\tilde{Z}^2 = \left |  \frac{1}{N}\sum_j \delta \theta_j \mathrm{e}^{i
    \theta_k} \right|^2\,,
\end{equation}
where the $k$ indices are randomly reshuffled versions of the original
$j$-indices. It turns out (see Fig.~\ref{fig:proj}a) that this reshuffling is sufficient to suppress
the logarithmic correction to scaling, i.e. that
\begin{equation}
\langle \tilde{Z}^2 \rangle \sim \frac{1}{N}  \; .
\end{equation}
In other words, the logarithmic correction is a manifestation of a
form of weak synchronization of the vector ${\bm
  \delta\theta}$, i.e. of non-trivial correlations between tangent and
phase space coordinates.  This can be also appreciated comparing 
the average square amplitude of the tangent vector components for
disordered (DU) and more regular (P-RE) frequency set-ups (again in
the discrete Kuramoto model). 
The results are plotted in Fig.~\ref{fig:proj}b, where we see that the
various time-averaged components, plotted as a function of the frequencies, are characterized
by substantially different amplitudes for the disordered case.  This is a major difference with respect to the regular setup, where
all components behave in the same way. 

Put in other words, our findings imply that in  disordered set-ups the LLE
tangent space vector localizes on a finite subset of frequencies,
thus inducing a logarithmic scaling correction in the fluctuations that
drive the tangent space dynamics and in the very LLE itself.
The presence of long-ranged order and/or of a finite minimum nearest
frequency gap, on the other hand, seems to delocalize the tangent
space vector, yielding simpler $1/N$ squared fluctuations in tangent space
dynamics and thus suppressing the logarithmic correction in the LLE.

So far we have developed and tested our theoretical argument
under a stochastic approximation (which is indeed rather crude, as it
neglects fluctuations correlations) and tested it in discrete-time
setups. It is desiderable to test a possible link between 
logarithmic corrections and localization properties of the corresponding Lyapunov vector in the
standard Kuramoto model.

Localization properties are captured by the so-called inverse
participation ratio (IPR),
\begin{equation}
Y_2 = \langle \sum_j \delta \theta_j^4\rangle_t
\end{equation}
where vector normalization ($\sum_j \delta \theta_j^2=1 $) is
understood and $\langle \cdot \rangle_t$ denotes a
time-average. Dealing with frequencies ensambles, of course, requires
a further average over different  realization IPRs, that is 
$Y_2 = \langle Y_2^{(\alpha)}\rangle_\alpha$.

In the thermodynamic limit, $Y_2 \to 0$ for delocalized vectors 
(actually $Y_2 \sim 1/N$ for perfectly extended structures). 
On the other hand, for localized vectors 
the IPR converges to a finite values, $Y_2 \to 1/\ell$,
where $\ell$ is the typical localization length. 

\begin{figure}[t!]
\includegraphics[scale=0.3]{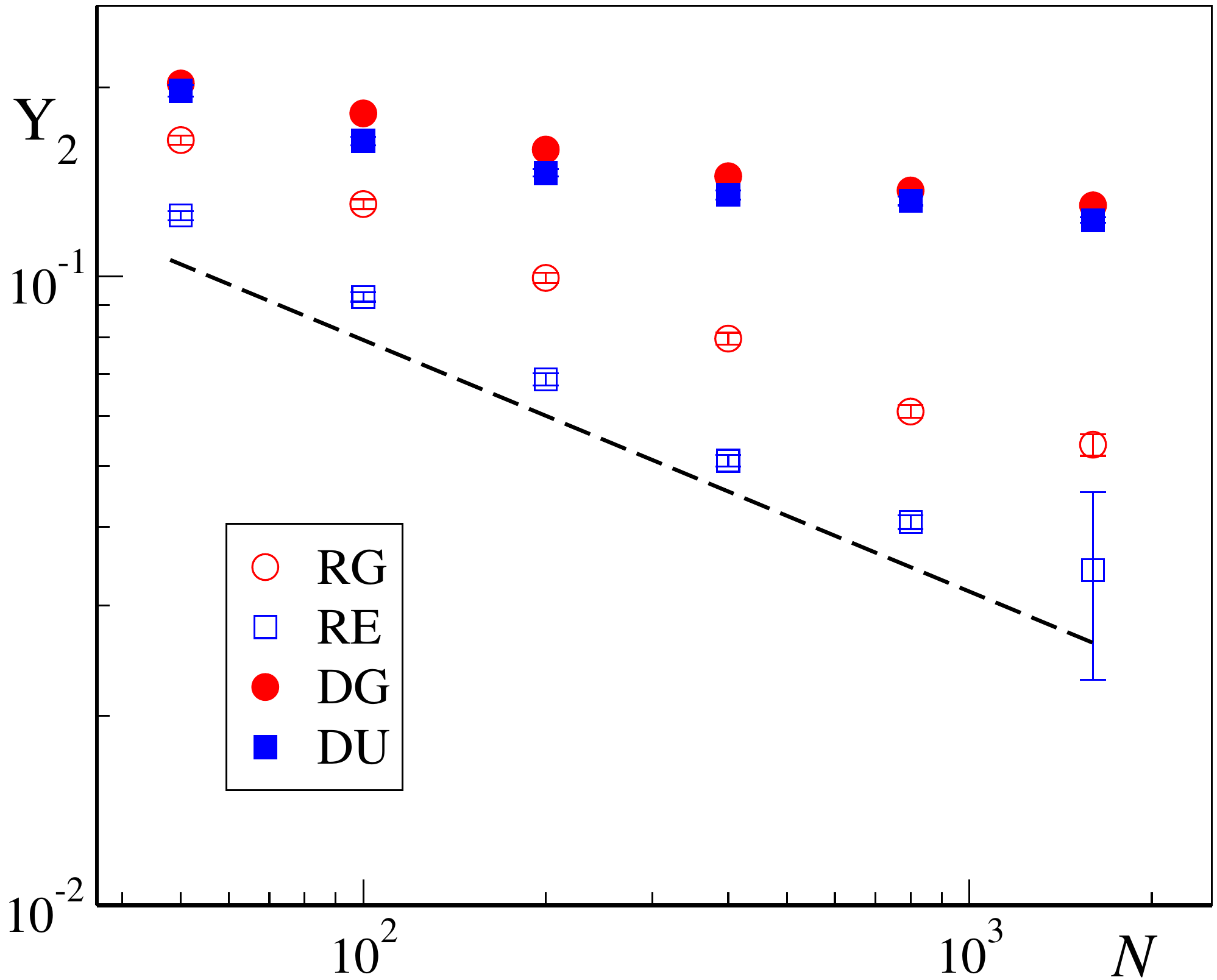}
\vspace{0.2cm}
\caption{Finite-size scaling of the tangent vector inverse participation ratio for the
  full Kuramoto model (\ref{Kuramoto}) and different frequency setups:
  Disordered Gaussian (DG, full red circles), Disordered Uniform (DU,
  full blue squares), Regular Gaussian RG, empty red circles) and
  Regular Equispaced (RE,
  empty blue squares). Axes are in a double-logarithmic
  scale. Simulation parameters as in Section
  \ref{sec:param}. Simulation parameters as in Section \ref{sec:param}). The
  dashed black line marks a power law decay $~ N^{-0.4}$.}
\label{figLast}
\end{figure}

We have computed the IPR for the full Kuramoto model (\ref{Kuramoto})
and the four frequency setups introduced in Section
\ref{sec:classes}. Our results, reported in Fig.~\ref{figLast},
unambigously show that disordered frequency choices (DG and DU) yield
a much more localized tangent vector, while regular ones (RG and RE) result in a
substantially extended structures.

\section{Conclusions}

We have discussed the way LLE scales to zero in systems of globally 
coupled oscillators in the asynchronous phase in the context of Kuramoto model. Our analysis
shows that the weak chaos in large but finite systems is essentially driven by 
fluctuations in the tangent-space dynamics. This is yet another instance of a rather 
general mechanims first unearthed in globally coupled chaotic units \cite{GCpap}, the difference
being that there the fluctuations, being an intrinsic property of the single oscillator dynamics,
stay finite in the thermodynamic limit. The same mechanism is active also in the Hamiltonian mean 
field, where it is able to support a finite Lyapunov exponent because of a series of subtle
phenomena~\cite{HMF}. 

Our numerical analysis reveals also that the scaling of the LLE depends on the way
the frequency distribution is built. In the physically meaningful and natural case of iid variables,
the LLE scales as $\lambda \sim \ln N /N$, while if the distribution is built using a
a regular protocol, the scaling is $\lambda \sim 1/N$ (this is, for instance the case
of equispaced frequencies first found in Ref.~\cite{Popovych}).
Given such differences, it is legitimate to ask whether the scaling results obtained
by using regular protocols are truly universal (see, e.g.~\cite{vanHemmen, Popovych, Pazo2005}
for the uniform distributions and \cite{Chate2015} for the analysis at criticality for a
normal distribution).

On a more mathematical, but fundamental level, it would be desirable to identify
the ultimate origin of the logarithmic correction.
It is natural to expect that the LLE depends on resonances between suitable
pairs or $n$-tuples of frequencies. In fact, the Lyapunov-vector structure 
reported in Fig.~\ref{figLast} indicates a pronounced localization on a special set
of frequencies.

This localization is probably the origin of the weak synchronization between tangent- and phase-space
dynamics signalled by the logarithmic correction in the scaling of $Z$ (\ref{eq:wa}).
However, the main questions are: how are the localization frequencies going to be selected?
How many are they?
Our simulations revealed the important role of quasi-degeneracies: 
they seem to be a necessary ingredient for the emergence of the logarithmic correction. 
However, this is cleary not enough as revealed by the two realizations of frequencies, 
characterized by the same spacing distributions, but
only one displaying the logarithmic correction. We speculate that long-range
order is a second important ingredient, but one cannot exclude that the ultimate
answer requires considering subtle number-theoretic properties of the frequency
spacings.

We limited ourselves to the analysis of the asynchronous phase. Preliminary numerical simulations,
suggest that the synchronous phase is characterized by a different scaling.
This is not entirely surprising, as the evolution in tangent space is different for two
reasons: (i) the Kuramoto order parameter has a finite value; (ii) a finite fraction of oscillators
are perfectly locked. 

Finally, notice that we have considered only the largest Lyapunov exponent, neglecting the
rest of the spectrum. Preliminary results indicate that only a handful of Lyapunov exponents (possibly a
non-extensive subset) are positive in finite systems~\cite{PikovPoliti}. 
A careful study of the full Lyapunov spectrum scaling is left for future work.

\begin{acknowledgments}
AP wishes to acknowledge D. Paz\'o for a preliminary analysis. 
All the authors thank H. Chat\'e for useful discussions.  
We acknowledge support from EU Marie Curie ITN grant n. 64256 (COSMOS).
\end{acknowledgments}

\appendix
\section{Frequency spacing distributions}
\label{app}
For completeness, we report here the analytical expression for the
frequency spacing distributions $\Pi(\Gamma)$studied in Section \ref{FSD}.
\begin{itemize}
\item
Poisson distribution,
\begin{equation}
\Pi(\Gamma)=\frac{1}{\Gamma_c} e^{-\Gamma/\Gamma_c}
\end{equation}
with $\Gamma_c \sim 1/N$ to ensure
the correct scaling.
This choice is fully equivalent to the disordered
uniformely distributed (DU) frequency distribution discussed in the main text.
\item
Inverse square root distribution,
\begin{equation}
\Pi(\Gamma)=\frac{1}{2 \sqrt{\Gamma \Gamma_c}}
\end{equation}
with $0<\Gamma<\Gamma_c$ and the cutoff $\Gamma_c \sim 1/N$ to ensure
the correct scaling.
\item
Uniform distribution,
\begin{equation}
\Pi(\Gamma)=\left\{
\begin{array}{c}
1\;\;\;\mbox{if}\;\; |\Gamma|\leq 1/2\\
0\;\;\;\mbox{if}\;\; |\Gamma|> 1/2
\end{array}
\right.\,.
\end{equation}
\item
Triangular distribution
\begin{equation}
\Pi(\Gamma)=\frac{2 \Gamma}{\Gamma_c^2}
\end{equation}
with $0<\Gamma<\Gamma_c$ and the cutoff $\Gamma_c \sim 1/N$\,.
\item
Pyramidal distributions
\begin{equation}
\Pi(\Gamma)=\left\{
\begin{array}{c}
C\left(\Gamma - \frac{1-2 a}{N-1}\right)
\;\;\;\mbox{if}\;\; \frac{1-2 a}{N-1}\leq\Gamma\leq\frac{1}{N}\\
C\left(\frac{1+2 a}{N-1}-\Gamma\right)
\;\;\;\mbox{if}\;\; \frac{1}{N}<\Gamma\leq \frac{1+2 a}{N-1}\\
\end{array}
\right.
\end{equation}
where $C=(N-1)^2/(2 a)$. The parameter $a \in [0,1/2]$ controls the frequency
spacing gap. For $a=1/2$ no finite gap is present, while for $a<1/2$ a
finite minimum gap appears (in the main text, $a=0.4$ for the gapped distribution).

\end{itemize}


\begin{thebibliography}{99}
\bibitem{NOTE2}
Without loss of generality $\bar{\omega}$ can be always set equal to 0,
as we are free to chose the frequency of the reference frame.
\bibitem{NOTE-GC}
In Ref. \cite{GCpap}, the diffusion constant is uncharacteristically
defined as the variance rather than as half the variance.
\end{thebibliography}
\end{document}